\documentclass[]{spie}  

\usepackage{amsmath,amsfonts,amssymb}
\usepackage{graphicx}
\usepackage{booktabs}
\usepackage[colorlinks=true, allcolors=blue]{hyperref}

\title{Coherent lidar for ride-hailing autonomous vehicles}

\author[a]{Alexander Y. Piggott}
\author[a]{Cathy Yunshan Jiang}
\author[a]{John Lam}
\author[a]{Blaise Gassend}
\author[a]{Simon Verghese}
\affil[a]{Waymo LLC, 680 E Middlefield Rd, Mountain View, CA 94043}

\authorinfo{Further author information: A.Y.P.: E-mail: apiggott@waymo.com}

\begin{document} 
\maketitle

\begin{abstract}
Coherent lidars promise a number of advantages over traditional time-of-flight lidars for autonomous vehicles. These include the direct measurement of target approach velocities via the Doppler effect, and near-immunity to interference from other lidars and sunlight. Furthermore, coherent lidars are compatible with a variety of solid-state beam steering technologies such as optical phased arrays, which may enable low-cost and compact lidars. In this manuscript, we discuss the headwinds facing the adoption of coherent lidar for autonomous ride-hailing vehicles and how they can be addressed. On the optics side, we explore how one can achieve the points per second and fields of view required for autonomous vehicles, and the resulting laser power requirements. We then discuss how these power levels could be achieved by co-packaging high-power semiconductor lasers and amplifiers with photonic integrated circuits, the preferred approach for low-cost coherent lidars. On the signal processing side, we discuss how to robustly disambiguate multiple returns in realistic environments and affordably meet the compute requirements.
\end{abstract}


\section{INTRODUCTION}

In contrast to traditional time-of-flight (ToF) lidar, which uses photodetectors to directly detect photons, coherent lidar detects the optical electric field using heterodyne receivers. Owing to the detection mechanism used by coherent lidar, we end up with a distinctly unique set of tradeoffs when compared to ToF lidar. For instance, coherent lidars measure velocity directly via the Doppler effect  \cite{jriemensberger_nature2020}, improving the segmentation of objects such as pedestrians in complex environments. They are also largely immune to interference from other lidars and sunlight \cite{jriemensberger_nature2020}. In addition, coherent lidars have a dynamic range advantage compared to time-of-flight lidars since they detect electric fields rather than optical power, and electric fields scale as the square root of power. Finally, since coherent lidars use single-mode continuous-wave lasers rather than the high peak-power multi-mode lasers typically used by ToF lidars, they are readily compatible with a wide range of solid-state beam steering technologies such as optical phased arrays \cite{cvpoulton_ieeejstqe2019}, Mach-Zehnder interferometer switches \cite{crogers_nature2021}, and MEMS waveguide switches \cite{xzhang_nature2022}, potentially enabling low-cost and compact lidars.

These advantages, however, do not come for free. Coherent lidars face a number of headwinds that have slowed down their adoption for autonomous ride-hailing vehicles:

\begin{enumerate}
\item To achieve the points per second and fields of view required by autonomous vehicles, coherent lidars must optimize their link transmission by using narrow, well focused beams, and "step-and-stare" beam scanning on the fast scan axis. This places stringent constraints on the beam steering mechanism and optical design.

\item Coherent lidars for these applications require aggregate optical powers on the order of several watts. Achieving these power levels at reasonable price points will require significant investments in light source integration.

\item Lidars operating in realistic environments will frequently detect multiple returns from the same laser beam. This can happen when the laser beam clips the edges of an object, or when it hits a partially transmissive target such as road spray or foliage. Naive ranging schemes for coherent lidar such as basic frequency-modulated continuous wave (FMCW) cannot disambiguate these multiple returns, necessitating the use of more complex modulation and post-processing schemes.

\item Coherent lidar has heavy compute requirements, necessitating the use of dedicated ASICs for affordable lidars. Coherent lidar ASICs are not readily available off the shelf, especially those that support the complex modulation schemes required for disambiguating multiple returns.
\end{enumerate}

In the following sections, we will discuss each of these challenges in greater detail, and explore how each of them can be overcome.

\section{COLLECTION EFFICIENCY OF COHERENT LIDAR}
\label{sec:collection_efficiency}
In this section, we discuss the physics of link transmission in coherent lidar. We then discuss the implications for the design of coherent lidars.

\begin{figure} [ht]
   \begin{center}
   \includegraphics[scale=0.75]{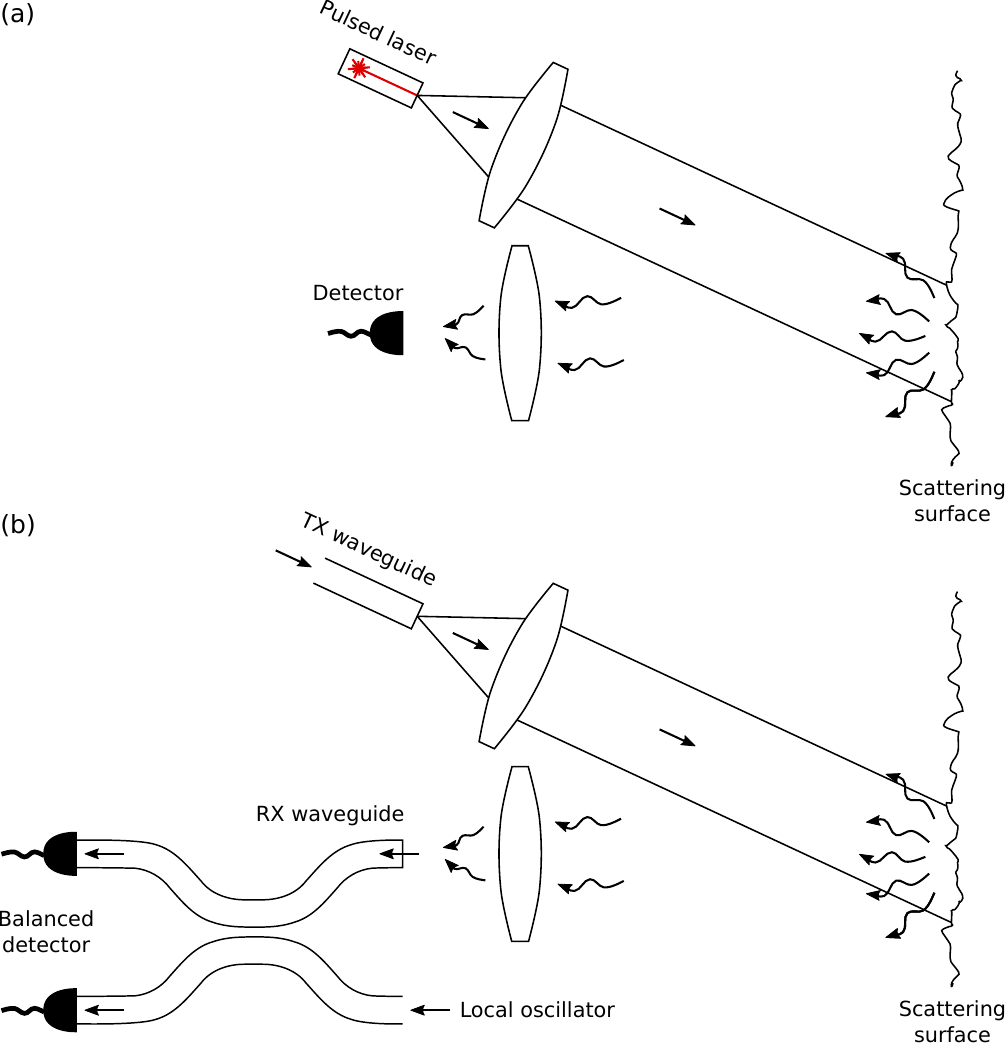}
   \end{center}
   \caption[example]{
        \label{fig:coherent_vs_tof_lidar}
        Lidar architectures. (a) Time-of-flight (ToF) lidar uses a pulsed laser on the transmit side and direct detection with a photodiode or other detector on the receiver side. (b) Coherent lidar uses a modulated single-mode laser as the transmitter. The receiver uses heterodyne detection, whereby scattered light is combined with a local oscillator on a beamsplitter before being directed onto a pair of photodiodes.
   }
   \end{figure}

Time-of-flight lidars use direct detection, as depicted in figure \ref{fig:coherent_vs_tof_lidar}(a). Here, scattered light from the target is collected by a lens and imaged onto a photodiode. As long as the image of the scattered beam falls within the detector footprint, all photons that pass through the receiver aperture are available for detection.

In contrast, coherent lidars use heterodyne detection. As depicted in figure \ref{fig:coherent_vs_tof_lidar}(b), received light is combined with a local oscillator beam before being directed onto a pair of photodiodes that constitute a balanced detector. As a result, a coherent lidar can \emph{only detect light in a single optical mode} per balanced detector. This is immediately obvious for a typical coherent lidar receiver constructed using single-mode fibers or on-chip single-mode waveguides. However, this is true regardless of how the coherent lidar is constructed, even if one were to use free-space beamsplitters and photodiodes \cite{aesiegman_ao1966}. As we will shall see, the link transmission of coherent lidars behaves differently from ToF lidars due to the single-mode nature of detection. In particular, within the Rayleigh range, which is where most automotive lidars will operate to optimize collection efficiency, link transmission is no longer proportional to the receiver aperture area or $1 / r^2$, where $r$ is the distance to the target.

\subsection{Uniformly illuminated target}
\begin{figure} [ht]
   \begin{center}
   \includegraphics[scale=0.75]{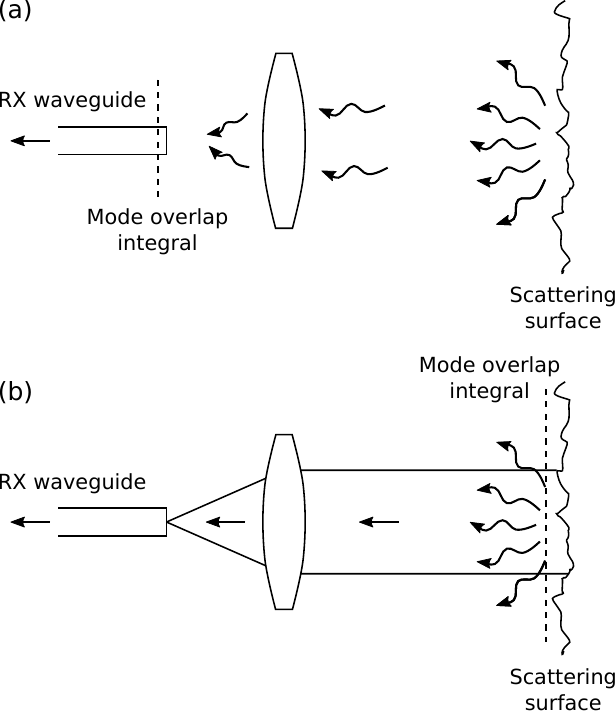}
   \end{center}
   \caption[example]{
        \label{fig:mode_overlaps}
        To calculate the power coupled into the receiver waveguide, we can either (a) take a mode overlap integral at the waveguide facet, or (b) back-propagate the receiver mode to the scattering surface, and take an overlap integral immediately above the scattering surface. The latter is more convenient for our purposes since it avoids propagating the complex and stochastic scattered fields through the receiver optics.
   }
   \end{figure}

To calculate the collection efficiency of a coherent lidar receiver, let us consider a receiver constructed using single-mode waveguides. As illustrated in figure \ref{fig:mode_overlaps}(a), we can obtain the power in the receiver waveguide by performing an overlap integral with the scattered optical field at the input facet of the waveguide. Alternatively, as illustrated in figure \ref{fig:mode_overlaps}(b), we can back-propagate the receiver mode to the scattering surface, and take an overlap integral immediately above the scattering surface. The latter is more convenient for our purposes, since it does not require us to propagate the complex scattered optical field through the receiver optics. Our goal, then, is to determine the expected power in the receiver mode immediately above the scattering surface.

   \begin{figure} [ht]
   \begin{center}
   \includegraphics[scale=0.75]{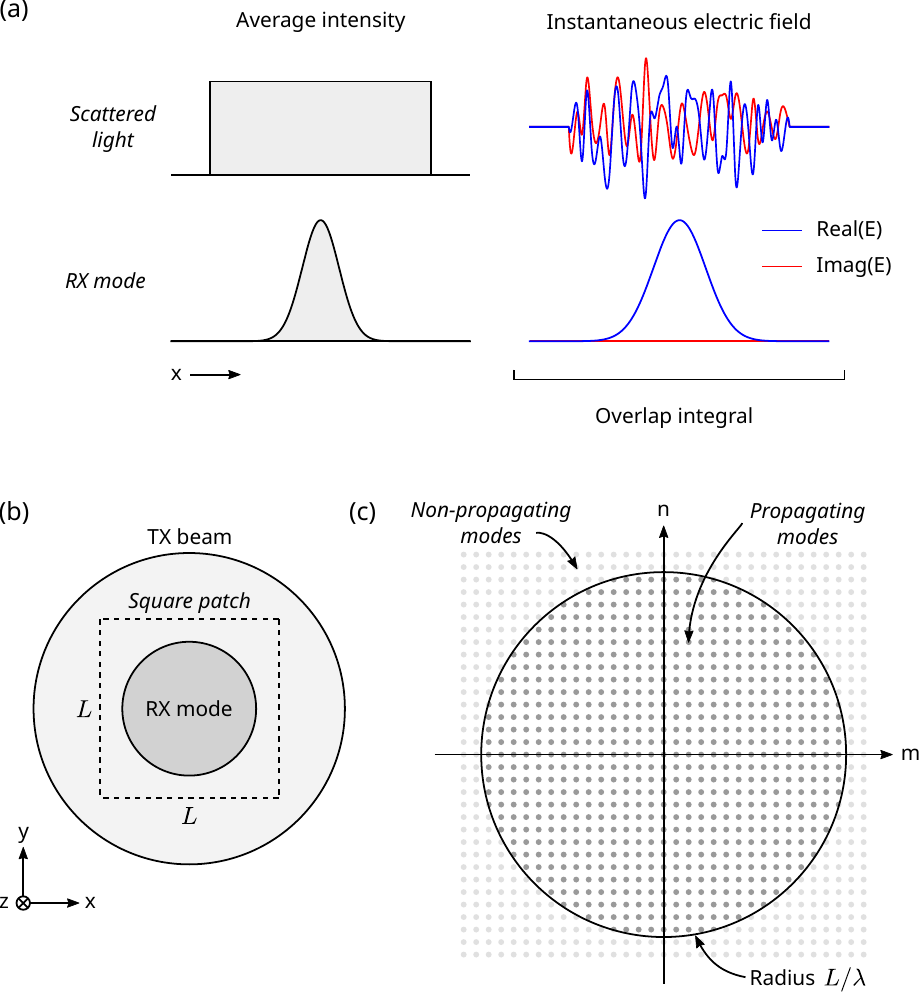}
   \end{center}
   \caption[example]{
        \label{fig:coupling_with_uniform_illum}
        Determining the expected power collected by a coherent lidar when the target is uniformly illuminated by the transmit beam. (a) The average intensity \emph{(left)} and instantaneous electric fields \emph{(right)} of the scattered light and receiver mode. The scattered light has uniform intensity across the entire receiver mode, but consists of a complex and stochastic electric field due to the rough scattering surface. The power in the receiver mode is given by the overlap integral of the scattered and receiver fields. (b) In our derivation, we consider a square patch immediately above the scattering surface that fully encloses the receiver mode, but in turn is smaller than the transmit beam. (c) There exists a mode for every pair of integers $m$ and $n$. The number of \emph{propagating} modes in the square patch is given by the number of modes within a circle of radius $L / \lambda$ in the $m - n$ plane. 
   }
   \end{figure} 

For simplicity, let us first consider the case where the transmit beam is much larger than receiver beam on the target, and the target is uniformly illuminated across the entire receiver beam, as illustrated in figure \ref{fig:coupling_with_uniform_illum}(a). We further assume that the target is a diffuse Lambertian surface, which will scatter light with equal probability into all available optical modes. Finally, we assume that the receiver optics are lossless, i.e. the lens and receiver waveguide have unity transmission.

To calculate the power collected by the receiver, consider a square patch of size $L \times L$ on the target surface. We choose the patch to be larger than the receiver beam, but smaller than the transmit beam, as shown in figure \ref{fig:coupling_with_uniform_illum}(b). Any optical field on this square patch can be expressed as a coherent superposition of the modes supported by this patch. These modes have electric fields $\mathbf{E}_{mn}(x, y, z)$ of the form
\begin{align}
\mathbf{E}_{nm}(x, y, 0) = \mathbf{E}_0 \exp\left(\frac{i 2 \pi m x}{L}\right) \exp\left(\frac{i 2 \pi n y}{L}\right),
\label{eqn:patch_mode_form}
\end{align}
where $m$ and $n$ are integers, and we have assumed the patch is in the $z = 0$ plane. Since the square patch is smaller than the transmit beam, and the Lambertian surface scatters light with equal probability into all available modes, we have the same expected power in all propagating modes supported by this patch. Furthermore, the receiver mode can be expressed as a coherent superposition of these modes, so the expected power $\left<P_\textit{RX}\right>$ collected by the receiver is the same as the expected power in any of the square patch modes.

We can find $\left<P_\textit{RX}\right>$ by dividing the scattered power $P_s$ in the patch by the number of propagating modes $N_\textit{modes}$ supported by the patch, i.e.
\begin{align}
\left<P_\textit{RX}\right> = \frac{P_s}{N_\textit{modes}}.
\label{eqn:power_counting_modes}
\end{align}
This can be rewritten in terms of the scattered light intensity $I_s$ and patch area $A$ as
\begin{align}
\left<P_\textit{RX}\right> = \frac{I_s A}{N_\textit{modes}}.
\label{eqn:intensity_counting_modes}
\end{align}
To find the number of propagating modes $N_\textit{modes}$, we note that for a propagating mode, the wavenumber in the $x-y$ plane must be smaller than the free-space wavenumber $k_0 = 2 \pi / \lambda$. Here, $\lambda$ is the wavelength. This condition can be written as
\begin{align}
k_0^2 \geq \frac{4 \pi^2}{L^2} \left(m^2 + n^2\right).
\end{align}
Rearranging using the definition of $k_0$ yields
\begin{align}
\frac{L^2}{\lambda^2} \geq  m^2 + n^2,
\end{align}
which describes a circle of radius $L / \lambda$ in the $m - n$ plane, as illustrated in figure \ref{fig:coupling_with_uniform_illum}(c). The number of modes in the square patch is therefore equal to the area of this circle, multiplied by a factor of two to account for polarization:
\begin{align}
N_\textit{modes} = \frac{2 \pi L^2}{\lambda^2}.
\end{align}
From equation \ref{eqn:intensity_counting_modes}, the expected power $\left<P_\textit{RX}\right>$ in the receiver mode is thus
\begin{align}
\left<P_\textit{RX}\right> = \frac{\lambda^2 I_s}{2 \pi},
\label{eqn:rx_power_uniform_beam}
\end{align}
where we have used $A = L^2$ as the area of the patch. A more rigorous derivation in [\citen{apiggott_arxiv2022}] yields the exact same result. The simple form of equation \ref{eqn:rx_power_uniform_beam} is quite surprising since it implies that received power \emph{depends only upon the intensity of the scattered light at the target}. The size of the receiver aperture and distance to the target do not matter as long as the receiver beam lands within the transmit beam.

\subsection{General case}
In the general case, we can no longer assume that the transmit beam is uniform across the target surface. For purpose of calculating the expected received power, it turns out that we can fully characterize the transmit and receive beams with their intensity distributions $I(x, y)$ at the target surface \cite{apiggott_arxiv2022}. For convenience, we define a beam's normalized intensity profile $\hat{I}(x, y)$ as
\begin{align}
\hat{I}(x, y) = \frac{I(x, y)}{\iint_\textit{target} I(x', y') \; dx' \, dy'},
\label{eqn:def_norm_intensity}
\end{align}
which satisfies the property
\begin{align}
1 = \iint_\textit{target} \hat{I}(x, y) \; dx \, dy.
\label{eqn:norm_intensity_has_unity_power}
\end{align}
In this formalism, the intensity $I_s(x, y)$ of the scattered light at the target surface can be written in terms of the Lambertian reflectance $R$, transmit power $P_\textit{TX}$, and normalized transmit beam profile $\hat{I}_\textit{TX}(x, y)$ as
\begin{align}
I_s(x, y) = R \, P_\textit{TX} \, \hat{I}_\textit{TX}(x, y).
\label{eqn:def_scattered_intensity}
\end{align}

If the transmit beam is non-uniform, we would expect the scattered light intensity term $I_s$ in equation \ref{eqn:rx_power_uniform_beam} to become a weighted average of the scattered intensity across the receiver beam. A detailed derivation in [\citen{apiggott_arxiv2022}] shows that $I_s(x, y)$ should be weighted by the normalized receive beam intensity $\hat{I}_\textit{RX}(x, y)$. More precisely, we should transform $I_s$ in equation \ref{eqn:rx_power_uniform_beam} as 
\begin{align}
I_s
\rightarrow
\iint_\textit{target} I_s(x, y) \; \hat{I}_\textit{RX}(x, y) \; dx \,dy.
\label{eqn:transform_scattered_intensity}
\end{align}
If we transform equation \ref{eqn:rx_power_uniform_beam} using equations \ref{eqn:def_scattered_intensity} and \ref{eqn:transform_scattered_intensity}, our expression for the expected received power becomes
\begin{align}
\left<P_\textit{RX}\right> = \frac{\lambda^2 \, R \, P_\textit{TX}}{2 \pi} \iint_\textit{target} \hat{I}_\textit{TX}(x, y) \; \hat{I}_\textit{RX}(x, y) \; dx \,dy.
\label{eqn:rx_power_nonuniform_case}
\end{align}
To account for lossy receiver optics, we can introduce the receiver transmission $\eta$, defined as the fraction of the backpropagated receiver mode that makes its way to the target. Any optical losses in the receiver chain, such as Fresnel reflections from lens surfaces or fiber facets, are captured by $\eta$. The received power in the most general case is therefore
\begin{align}
\left<P_\textit{RX}\right> = \frac{\lambda^2 \, \eta \, R \, P_\textit{TX}}{2 \pi} \iint_\textit{target} \hat{I}_\textit{TX}(x, y) \; \hat{I}_\textit{RX}(x, y) \; dx \,dy.
\label{eqn:rx_power_general_case}
\end{align}

\subsection{Monostatic case}
Many coherent lidars employ a monostatic architecture to maximize the overlap integral in equation \ref{eqn:rx_power_general_case} and therefore the collection efficiency. In the monostatic case, the transmitted and received light share the same optical path, so $\hat{I}_\textit{TX}(x, y) = \hat{I}_\textit{RX}(x, y) = \hat{I}(x, y)$, and we can define an effective beam area at the target,
\begin{align}
\frac{1}{A_\textit{eff}} = \iint_\textit{target} \hat{I}^2(x, y) \; dx \, dy.
\label{eqn:def_eff_beam_area}
\end{align}
The effective beam areas of some common beam profiles are listed below. For these examples, the effective beam area closely matches what one would intuitively expect.
\begin{center}
\begin{tabular}{ c c c }
  \toprule
  Type & Intensity $I(x, y)$ & Effective area ($A_\mathit{eff}$) \\
  \midrule
  Flat-top
    & $\begin{cases}
        1, & x^2 + y^2 < a^2 \\
        0, & \mathrm{otherwise.}
      \end{cases}$
    & $\pi a^2$ \vspace{0.2cm} \\ 
   Gaussian 
    & $\exp\left( - \dfrac{2 (x^2 + y^2)}{w^2} \right)$
    & $\pi w^2$ \vspace{0.2cm} \\
  \bottomrule
\end{tabular}
\end{center}
The received power can then rewritten in terms of the beam area $A_\textit{eff}$ as
\begin{align}
\left<P_\textit{RX}\right> = \frac{\lambda^2 \, \eta \, R \, P_\textit{TX}}{2 \pi A_\textit{eff} }.
\label{eqn:rx_power_monostatic}
\end{align}
This implies that the link transmission of a monostatic coherent lidar depends only upon the beam area at the target $A_\textit{eff}$, target reflectance $R$, and transmission efficiency $\eta$ of the receiver chain. To optimize the collection efficiency of a coherent lidar, the optical designer should therefore minimize the worst-case beam size across all ranges the lidar is expected to operate at.

\subsection{Experimental validation}
To validate our coherent lidar link transmission model, we performed detailed measurements using a homebuilt coherent lidar operating at $1550~\text{nm}$. The lidar was a monostatic frequency-modulated continuous-wave (FMCW) system using a tunable diode laser as the light source. Additional details on the lidar setup are provided in Appendix \ref{sec:tlafmcw_setup}.

   \begin{figure} [ht]
   \begin{center}
   \includegraphics[scale=0.75]{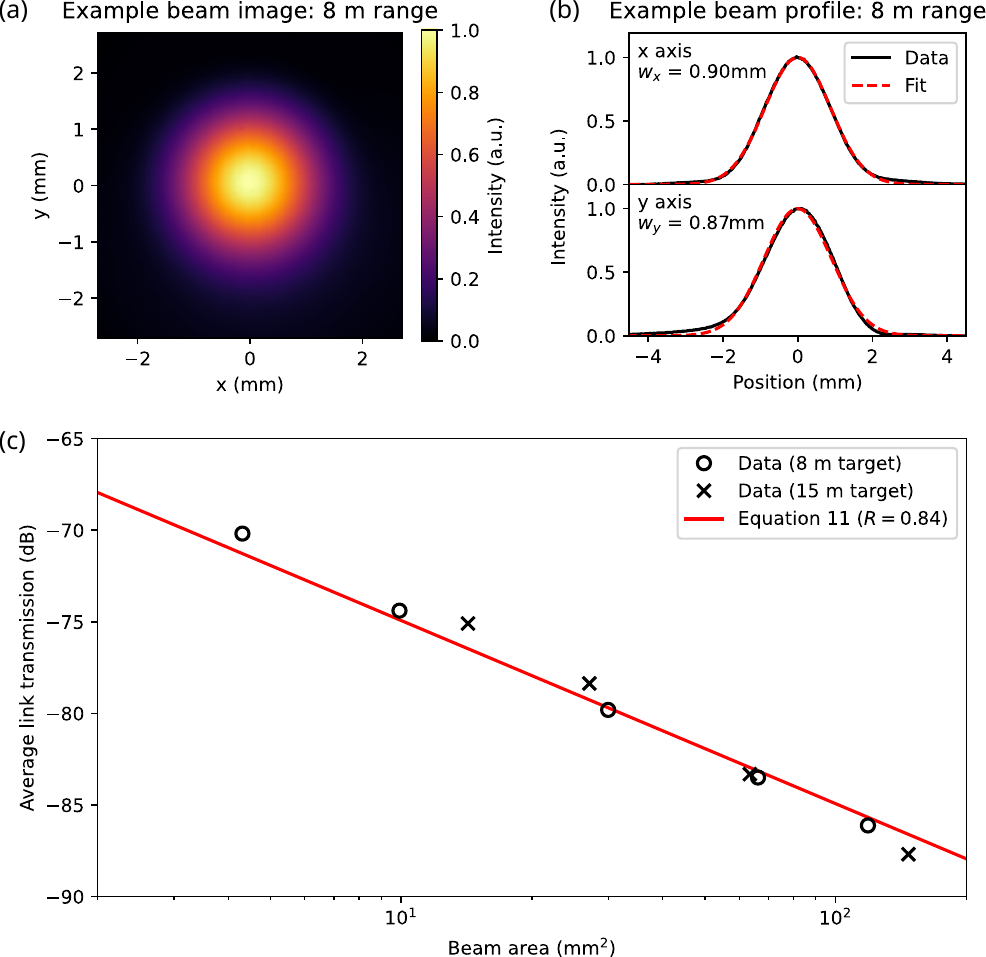}
   \end{center}
   \caption[example]{
        \label{fig:link_transmission_measurements}
        Experimental validation of the coherent lidar link transmission model. (a) The beam image at the target was directly captured by placing an AlliedVision G-032 InGaAs camera sensor in the beam path. (b) Due to the highly Gaussian nature of the beams, the beam area was estimated by fitting a 2D Gaussian to the measured beam profile. (c) The measured link transmission closely matches equation \ref{eqn:rx_power_monostatic} for the link transmission of a monostatic coherent lidar. For the model, we used an independently measured value of 84\% for the target material (Labsphere Permaflect 80) retroreflectance. The link transmission depends only upon the beam area at the target, and does not change with distance to the target.
   }
   \end{figure} 

Link transmission was measured as a function of effective beam area for a Labsphere Permaflect 80 target placed at $8~\text{m}$ and $15~\text{m}$. Beam images at the target were directly captured by placing an infrared camera sensor (AlliedVision G-032 InGaAs sensor) in the beam path, as illustrated in figure \ref{fig:link_transmission_measurements}(a). A single-mode fiber with a collimating lens was used for the transmit / receive path, resulting in an almost perfectly Gaussian beam profile. As illustrated in figure \ref{fig:link_transmission_measurements}(b), this allowed us to estimate the beam area by fitting a 2D Gaussian to the measured beam profile. The beam area at the target was swept by adjusting the focus of the collimating lens; no other changes were made to the optics between measurements. To eliminate noise from speckle effects, the beam was scanned across the target using a galvo mirror, and 1000 lidar measurements were averaged for each data point. Additional details on the measurements are provided in Appendix \ref{sec:link_measurement}.

Since the lidar used in this experiment has a monostatic configuration, the correct reflectance $R$ to use in equation \ref{eqn:rx_power_monostatic} is the \emph{retroreflectance}, defined as the bidirectional reflectance distribution function (BRDF) relative to an ideal Lambertian for identical incidence and reflectance angles. It has widely been observed that for most diffuse materials, there is an increase in reflectance as the angle between the incident and scattered rays goes to zero \cite{bhapke_jgr1963, poetking_jgr1966, ykuga_jpsaa1984, ltsang_josaa1984, bwhapke_science1993}. In internal testing at Waymo, we have found that Labsphere Permaflect material also exhibits this spike in retroreflectance. We thus independently measured the retroreflectance of the target using an in-house retroreflectance setup, and obtained a value of 84\%. This is slightly higher than the Lambertian reflectance value of 77\% provided by the manufacturer's datasheet \cite{permaflect80_labsphere2017} at a wavelength of $1550~\mathrm{nm}$.

As shown in figure \ref{fig:link_transmission_measurements}(c), our measurements closely match the values predicted by equation \ref{eqn:rx_power_monostatic} for the link transmission of a monostatic coherent lidar. As expected, the link transmission depends only upon the effective beam area, and is independent of the range to the target. 

\subsection{Implications for the design of coherent lidars}
As discussed earlier, equation \ref{eqn:rx_power_monostatic} implies that coherent lidars should seek to minimize the beam area of the transmit and receive beams across all ranges the lidar is operated at. This is best achieved by using well-collimated, diffraction limited beams in a monostatic configuration. Collimated Gaussian beams are nearly optimal in terms of maximizing link transmission since they minimize beam divergence for a given waist diameter.

   \begin{figure} [ht]
   \begin{center}
   \includegraphics[scale=0.75]{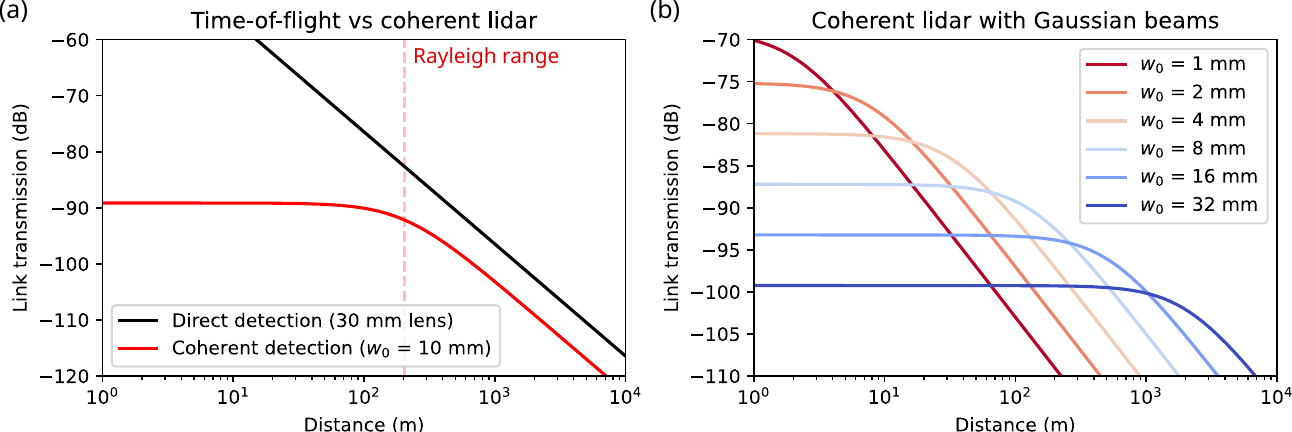}
   \end{center}
   \caption[example]{
        \label{fig:coherent_link_transmission}
        Lidar link transmission for a 100\% Lambertian target. (a) Comparison between the link transmission of a time-of-flight lidar and a monostatic coherent lidar. We assume the coherent lidar is operating at $1550~\text{nm}$, and using a collimated Gaussian beam with a $1/e^2$ radius $w_0$ of $10~\text{mm}$. For the time-of-flight lidar, we use a lens aperture that is just large enough to avoid significant clipping of the coherent lidar's Gaussian beam. (b) Link transmission for coherent lidars using collimated Gaussian beams spanning a range of beam widths. Larger beams trade short-range transmission for better transmission at long ranges.
   }
   \end{figure}

In figure \ref{fig:coherent_link_transmission}(a), we compare the link transmission of a monostatic coherent lidar using a collimated Gaussian beam, and a time-of-flight lidar with a comparable aperture size. Within the Rayleigh range of the Gaussian beam, the beam area and therefore link transmission of the coherent lidar is essentially flat. At longer ranges, the coherent lidar link transmission rolls off as $1/r^2$, tracking the link transmission of the time-of-flight lidar. At these longer ranges, the link transmission of the coherent lidar is approximately $6~\text{dB}$ lower than the time-of-flight lidar, of which $3~\text{dB}$ is due to the fact that the coherent lidar only detects light in one polarization. The remaining $\sim 3~\text{dB}$ is due to the larger effective aperture of the time-of-flight lidar: the lens aperture for the time-of-flight lidar was chosen to be large enough to avoid heavily clipping the wings of the Gaussian beam, but the coherent lidar mostly collects light from the center of the Gaussian beam.

In figure \ref{fig:coherent_link_transmission}(b), we illustrate the impact of waist size on the link transmission of coherent lidars using collimated Gaussian beams. Beams with larger waists have larger beam areas at short range, but diverge less due to diffraction and therefore have smaller beam areas at long range. This highlights the fundamental tension between short and long-range sensitivity in coherent lidars: improving link transmission at long ranges necessarily requires sacrificing link transmission at short range. In some ways, however, this is convenient for the lidar designer: the flat link transmission as a function of range reduces the dynamic range that the lidar receiver must support.

   \begin{figure} [ht]
   \begin{center}
   \includegraphics[scale=0.75]{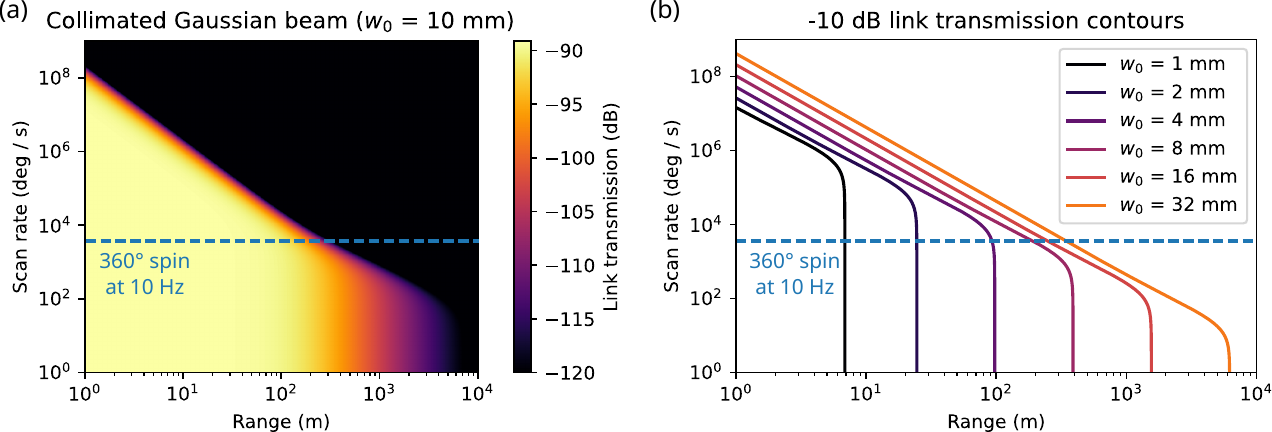}
   \end{center}
   \caption[example]{
        \label{fig:scanning_link_loss}
        Link transmission as a function of angular scan rate for a 100\% Lambertian target. The dashed blue lines indicate the scan rate for a lidar spinning $360^\circ$ at $10~\text{Hz}$, a common lidar architecture for autonomous vehicles. (a) Link transmission for a monostatic coherent lidar operating at $1550~\text{nm}$ and using a collimated Gaussian beam with a $1/e^2$ beam radius $w_0$ of $10~\text{mm}$. At any given range, the transmission falls off very quickly after reaching a critical angular scan rate. (b) We can assess the regime over which a coherent lidar will perform well by plotting the contour over which the link transmission drops by $10~\text{dB}$ below its zero-distance transmission. Here, we plot these contours for monostatic coherent lidars using collimated Gaussian beams with a range of beam widths. 
   }
   \end{figure}

The requirement for narrow, collimated beams also limits the angular scan rate of coherent lidars. Due to the finite speed of light, the transmit and receive beams effectively look in different directions when the lidar is scanned across the scene. For a given angular scan rate $\omega$ in $\text{radians} / \text{s}$, the physical offset $\Delta x$ between the transmit and receive beams on a target at a distance of $r$ is
\begin{align}
\Delta x \approx \frac{2 \, r^2 \, \omega}{c}.
\label{eqn:beam_scanning_separation}
\end{align}
Here, $c$ is the speed of light. The coherent lidar link transmission can then be calculated by applying equation \ref{eqn:rx_power_general_case} to the laterally offset transmit and receive beam profiles at the target. Due to the small beams required by coherent lidars to maximize link transmission, they tend to be more sensitive to these offsets than typical time-of-flight lidars.

The link transmission for a typical coherent lidar as a function of range and scan rate is illustrated in figure \ref{fig:scanning_link_loss}(a). For any given range, the link transmission falls off abruptly after reaching a critical scan rate. To illustrate the impact of beam diameter, the contour over which link transmission drops $10~\text{dB}$ below the zero-distance link transmission is plotted in figure \ref{fig:scanning_link_loss}(b) for a variety of beam sizes. Beams with wider waists are slightly more robust to beam scanning. However, regardless of beam size, it is clear that spinning the entire lidar $360^\circ$ at $10~\text{Hz}$, a common architecture for time-of-flight lidars on autonomous vehicles, is impractical for target ranges much above $100~\text{m}$. Furthermore, we can see that mechanically scanning the beam in the fast scan axis is impractical for all but the shortest ranges. "Step-and-stare" beam scanning, where the lidar beams dwell at a fixed angle for an entire acquisition before jumping to the next angle, is generally preferred. This is possible using a variety of solid-state beam steering approaches such as optical phased arrays \cite{cvpoulton_ieeejstqe2019} and focal plane arrays \cite{xzhang_nature2022, crogers_nature2021}.

\section{COHERENT LIDAR DETECTION STATISTICS}
\label{sec:detection_statistics}
Coherent lidars are reasonably sensitive, and are often referred to as having ``near single-photon sensitivity''. In this section, we quantify this statement, deriving the number of photons required for the robust detection of a target under a variety of conditions. This, in turn, will allow us to compute the required transmit power of coherent lidars.

\subsection{Carrier-to-noise ratio}
Coherent lidars are typically operated in the shot-noise limited regime, where the shot noise of the local oscillator light dominates over all other noise sources in the receiver chain. In this regime, the \emph{carrier-to-noise ratio}, defined as the expected signal energy divided by the expected noise energy, is equal to the expected number of detected photons $N$ \cite{hpyuen_ol1983, mjcollet_jmo1987, pgatt_spielrta2001, apiggott_arxiv2022}. This is the basis for the sensitivity of coherent lidars.

\begin{figure} [ht]
   \begin{center}
   \includegraphics[scale=0.75]{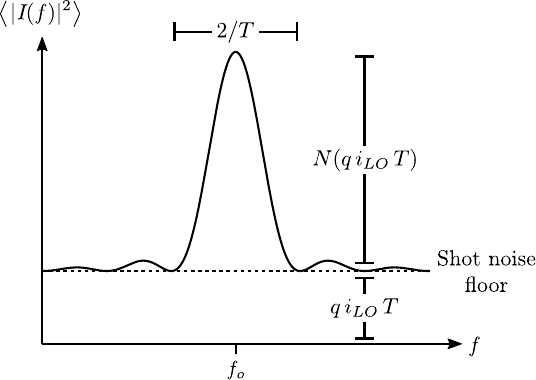}
   \end{center}
   \caption[example]{
        \label{fig:cnr_of_coherent_lidar}
        The average energy spectral density of a frequency-modulated continuous wave (FMCW) lidar operating in the shot noise limited regime. The carrier-to-noise ratio (CNR) is equal to the average number of received photons $N$, assuming we use a rectangular FFT window. The CNR will be slightly lower for other choices of window.
   }
   \end{figure}

As a concrete example, let us consider the frequency-modulated continuous-wave (FMCW) lidar. Here, the received signal is at some frequency offset $f_o$ from the local oscillator laser, producing a beat note in the photocurrent at the offset frequency (see section \ref{sec:disambiguating_returns} for additional details). The photocurrent coming out of the balanced detector $i(t)$ is given by the sum of the beat note or ``carrier'' photocurrent $c(t)$ and shot noise $s(t)$,
\begin{align}
i(t) = c(t) + s(t).
\end{align}
The carrier photocurrent is a sinusoid with a frequency of $f_o$,
\begin{align}
c(t) = 2 \sqrt{i_\textit{LO} \, i_\textit{sig}} \cos\left( 2 \pi f_o t \right),
\end{align}
where $i_\textit{LO}$ and $i_\textit{sig}$ are the time-averaged local oscillator and signal photocurrents \cite{apiggott_arxiv2022}. Typically, an FMCW lidar will find the carrier tone by digitizing $i(t)$, taking a Fourier transform over some time window of length $T$, and looking for a peak in the spectrum $I(f)$. The expected energy spectral density of the photocurrent $\left< |I(f)|^2 \right>$ is given by the sum of the carrier $|C(f)|^2$ and shot noise $\left< |S(f)|^2 \right>$ energy spectral densities,
\begin{align}
\left< |I(f)|^2 \right> = |C(f)|^2 + \left< |S(f)|^2 \right>.
\end{align}
One can show \cite{apiggott_arxiv2022} that if we use a rectangular FFT window, the carrier energy spectral density at the beat frequency $f_o$ is
\begin{align}
|C(f_o)|^2 = N \, q \, i_\textit{LO} \, T,
\label{eqn:carrier_energy}
\end{align}
where $q$ is the elementary charge, and $N$ is the average number of detected photoelectrons over time $T$, i.e. the average number of received photons multiplied by the quantum efficiency of the balanced detector. Meanwhile, the expected energy spectral density of the shot noise is
\begin{align}
\left< |S(f)|^2 \right> =  q \, i_\textit{LO} \, T.
\label{eqn:shot_noise_energy}
\end{align}
The expected spectrum $\left< |I(f)|^2 \right>$ of the photocurrent is illustrated in figure \ref{fig:cnr_of_coherent_lidar}.

The carrier-to-noise ratio (CNR) is given by the ratio of the carrier $|C(f_o)|^2$ and shot noise $\left< |S(f_o)|^2 \right>$ energy spectral densities at the beat frequency $f_o$. Using equations \ref{eqn:carrier_energy} and \ref{eqn:shot_noise_energy}, we therefore find that
\begin{align}
\text{CNR} = \frac{|C(f_o)|^2}{\left< |S(f_o)|^2 \right>} = N.
\label{eqn:coherent_lidar_cnr}
\end{align}
In other words, the CNR is equal to the average number of detected photoelectrons $N$.

\subsection{Required number of photons for robust detection}
\label{subsec:detection_sensitivity}
To study the statistics of coherent lidar detection, we continue with our concrete example of the FMCW lidar. Typically, the received signal in an FMCW lidar is digitized and fed through a fast Fourier transform (FFT) to generate a spectrum. The complex value of the FFT output in the $i$th frequency bin, denoted as $V_i$, is the sum of the carrier amplitude $C_i$ and shot noise $S_i$,
\begin{align}
V_i = C_i + S_i.
\end{align}
Due to speckle, the carrier amplitude $C_i$ is a complex Gaussian random variable \cite{jwgoodman_chapter1975}. As discussed in the previous section, the expected energy of $C_i$ is equal to the expected number of detected photoelectrons $N$,
\begin{align}
\left< |C_i|^2\right> = N~\text{photoelectrons}.
\end{align}
Similarly, since shot noise is additive white Gaussian noise, the shot noise term $S_i$ is a complex Gaussian random variable with an expected energy of 1 photoelectron,
\begin{align}
\left< |S_i|^2\right> = 1~\text{photoelectron}.
\end{align}
The total complex value $V_i$ is therefore also a complex Gaussian random variable with expected energy
\begin{align}
\left< |V_i|^2\right> = \left< |C_i + S_i|^2\right> = \left< |C_i|^2\right> + \left< |S_i|^2\right> = N + 1~\text{photoelectrons},
\end{align}
since the shot noise $S_i$ and carrier amplitude $C_i$ are independent zero-mean random variables. These statistics generalize to the majority of coherent lidar detection schemes, including phase-code modulation schemes \cite{pgatt_spielrta2001}.

\begin{figure} [ht]
  \begin{center}
  \includegraphics[scale=0.75]{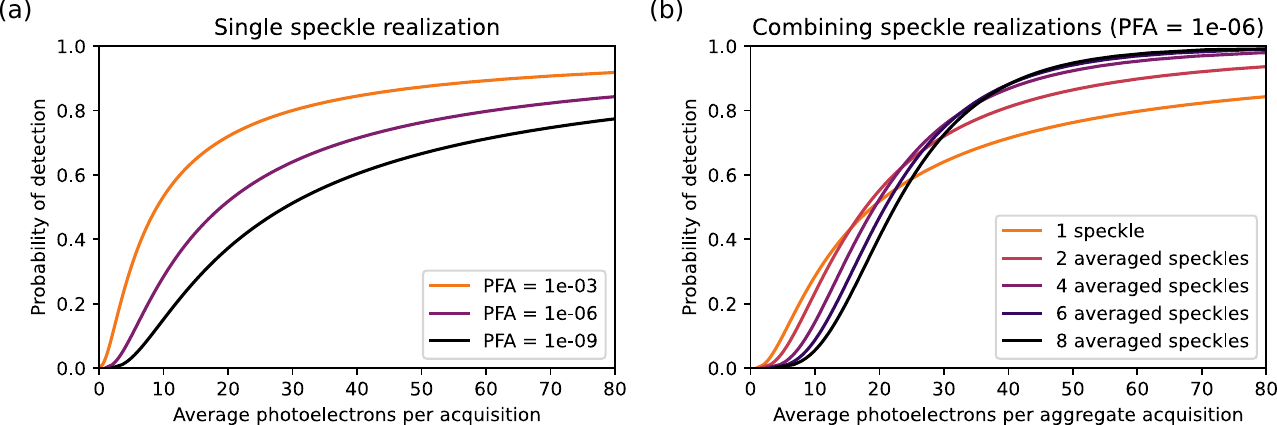}
  \end{center}
  \caption[example]{
        \label{fig:coherent_detection_statistics}
        Detection probability as a function of the number of detected photoelectrons, and the probability-of-false-alarm (PFA) per distance bin. Automotive lidars typically target PFA $\leq 10^{-6}$ per distance bin. (a) If we only use a single speckle realization, very high average numbers of photoelectrons are required to reach $> 90\%$ detection probability. (b) If we incoherently combine $n$ speckle realizations into a single aggregate acquisition, for example by summing the frequency spectra of several FMCW lidar acquisitions, significantly fewer photoelectrons are required to reach $> 90\%$ detection probability. 
  }
  \end{figure}

To detect returns, we take the absolute square of the complex value $V_i$ to yield the signal energy $E_i$,
\begin{align}
E_i = |V_i|^2,
\end{align}
and apply thresholding. The energy $E_i$ is an exponential random variable with a mean value of $N + 1$, and a probability density function $f_{E_i}(x)$ given by
\begin{align}
f_{E_i}(x) = \frac{e^{-x / \left(N + 1\right)}}{N + 1}.
\end{align}
For a given detection threshold $t$, the probability of detection $P_D$ is the probability that $E_i > t$. It is straightforward to show that
\begin{align}
P_\textit{D} = \text{P}(E_i > t) = e^{- t / \left(N + 1\right)}.
\label{eqn:pd_single_speckle}
\end{align}
The detection threshold is generally set to achieve a specific probability of false alarm $P_\textit{FA}$, which is the probability that $E_i > t$ if we happen to have zero detected photoelectrons. Setting $N = 0$ in equation \ref{eqn:pd_single_speckle}, we find that the probability of false alarm is 
\begin{align}
P_\textit{FA} = e^{-t}.
\label{eqn:pfa_single_speckle}
\end{align}
The detection probability as a function of the expected number of detected photoelectrons is plotted in figure \ref{fig:coherent_detection_statistics}(a), with separate traces for different probabilities of false alarm. Automotive lidars typically target probabilities of false alarm $\leq 10^{-6}$ per distance bin, or equivalently for FMCW lidar, per frequency bin. Only 10 to 30 photoelectrons are required to reach a detection probability of $50\%$, depending upon the desired PFA. Due to the statistics of speckle, however, very high average numbers of photoelectrons are required to reach $> 90\%$ detection probability. High detection probabilities are generally desirable in automotive lidars, especially for targets that are not at the detection limits of the lidar.

One approach for reducing the average number of photoelectrons required to reach high detection probabilities is to sum up the signal from several independent speckle realizations. This can be done by scanning the lidar beam across the scene, which will cause the carrier amplitude to decohere \cite{jhshapiro_ao1985, jywang_ao1986, esbaumann_ol2014} over successive acquisitions. Let us consider what happens if we combine the signal energy from $n$ acquisitions into a single aggregate energy $A_i$. If we denote the signal energy of the $j$th acquisition as $E_i^{(j)}$, the aggregate energy is given by
\begin{align}
A_i = \sum_{j = 1}^{n} E_i^{(j)}.
\end{align}
Assuming the speckle realizations of the acquisitions are fully independent, $A_i$ is the sum of $n$ independent exponential random variables, each with a mean of $N + 1$. The sum of identical exponential random variables follows the Erlang distribution \cite{rdyates_book2014}, whose probability density function is
\begin{align}
f_\textit{Erlang}(x; \, k, \beta) = \frac{x^{k - 1} e^{-k/\beta}}{\beta^k (k - 1)!}.
\end{align}
Here, $k$ is the number of random variables being summed, and $\beta$ is their mean value. We can denote the cumulative distributive function (CDF) of the Erlang distribution as $F_\textit{Erlang}(x; \, k, \beta)$, which is is typically evaluated numerically. In the case of the aggregate energy $A_i$, we can identify $k = n$ and $\beta = N + 1$. The probability of detection $P_D$ for a detection threshold of $t$ can therefore be given in terms of the CDF of the Erlang distribution as
\begin{align}
P_D = F_\textit{Erlang}(t; \, n, N + 1).
\end{align}
Similarly, the probability of false alarm $P_\textit{FA}$ is given by the probability of crossing the detection threshold given $N = 0$, i.e.
\begin{align}
P_\textit{FA} = F_\textit{Erlang}(t; \, n, 1).
\end{align}
Figure \ref{fig:coherent_detection_statistics}(b) shows the detection probability for various numbers of averaged speckle realizations $n$. On the x-axis, we have plotted $n \times N$, the total number of photoelectrons expected across all $n$ acquisitions. Significantly fewer detected photoelectrons are required to reach high detection probabilities, albeit with diminishing returns for $n > 4$.

\section{LASER SOURCE REQUIREMENTS}
The theoretical limits of coherent lidar sensitivity discussed in \ref{sec:detection_statistics} can only be achieved if the laser noise meets stringent requirements, as outlined below.
\begin{enumerate}
  \item The coherence time of the laser must be at least 5 to 10 times greater than the round trip travel time to avoid noticeably broadening the beat tone between the received light and local oscillator \cite{lrichter_ieeejqe1986, lbmercer_jlt1991}, which would reduce the signal-to-noise ratio. For a coherent lidar with an maximum range of $250~\text{m}$, this corresponds to a linewidth of approximately $60 - 120~\text{kHz}$.
  \item The relative intensity noise (RIN) of the laser must be low enough that shot noise dominates over local oscillator intensity noise in the detectors. Typical balanced detectors have common-mode rejection ratios (CMRR) on the order of $20 - 40~\text{dB}$, which only modestly relaxes the RIN requirements.
  \item The phase noise skirt of the laser determines the dynamic range of the lidar. In the case of FMCW lidar, this is affects returns that are close together in range \cite{svenkatesh_jlwt1993}, whereas in phase coded lidar, the dynamic range is affected for all ranges. Real-world lidar scenes have tremendous dynamic range: a retroreflector may be more than $1000 \times$ brighter than an ideal Lambertian, whereas a wet dark car may be dimmer than a 1\% Lambertian, corresponding to a dynamic range of more than $50~\text{dB}$ in reflectance. Phase noise skirts are particularly critical for monostatic lidars, which typically have significant zero-range returns from backreflections in the transmit / receive optics. The minimum sensing range of a monostatic FMCW lidar is largely determined by the phase noise skirt of the laser.
\end{enumerate}

Today, these laser noise requirements are readily achieved by diode lasers designed for coherent lidar applications. For example, several vendors provide distributed-feedback (DFBs) lasers with linewidths in the low tens of kHz. Significantly more challenging, however, are the laser power requirements. To illustrate this, we will perform a case study of an advanced driver assistance system (ADAS) lidar.

\subsection{Case study: a typical long-range ADAS lidar}
\label{subsec:adas_lidar}

\begin{table}[ht]
\begin{center}
\begin{tabular}{ l c }
  \toprule
  \textbf{Parameter} & \textbf{Value} \\
  \midrule
  System requirements \\
  \qquad Range & $250~\text{m}$ \\
  \qquad Field of view & $120^\circ\text{H} \times 25^\circ\text{V}$ \\
  \qquad Resolution & $0.05^\circ$ \\
  \qquad Target reflectance &  5\% \\
  \qquad Probability of detection & 90\% \\
  \qquad Probability of false alarm & $10^{-6}$ per range bin \\
  \midrule
  Design choices \\
  \qquad Wavelength & $1550~\text{nm}$ \\
  \qquad Gaussian beam waist $(w_0)$ & $10~\text{mm}$ \\
  \qquad Photodiode quantum efficiency & 90\% \\
  \qquad Chirps per point & 2 \\
  \qquad Number of averaged speckles & 4 \\
  \midrule
  Derived quantities \\
  \qquad Shots per second & $1.2 \times 10^7$ \\
  \qquad Photons per detection & 50 \\
  \qquad Worst-case link transmission & $-106.1~\text{dB}$ \\
  \qquad TX energy per shot & $0.52~\mu\text{J}$ \\
  \qquad TX power & $6.2~\text{W}$ \\
  \bottomrule
\end{tabular}
\end{center}
\caption{
\label{table:adas_lidar}
Parameters for an idealized FMCW lidar that meets typical advanced driver assistance system (ADAS) lidar requirements. To provide a hard lower bound on the laser power requirements, we have assumed a monostatic coherent lidar which uses lossless optics, collimated diffraction-limited Gaussian beams, a circulator to separate transmitted and received light, and a shot-noise limited receiver.
}
\end{table}

System level parameters for an FMCW lidar that meets current ADAS requirements are provided in table \ref{table:adas_lidar}. These requirements are typical of what automotive manufacturers are requesting, and what time-of-flight lidar vendors can currently provide. To provide a lower bound on the laser power requirements, we have assumed an ideal monostatic coherent lidar which uses lossless optics, collimated diffraction-limited Gaussian beams, a circulator to separate transmitted and received light, and a shot-noise limited receiver.

Due to the high shots-per-second requirement, this idealized FMCW lidar requires an aggregate transmit power of $\sim6~\text{W}$. However, circulators are expensive optical components that are unlikely to be used in vehicle lidars in the near future, and eliminating the circulator incurs an optical loss of $3~\text{dB}$. Realistic implementations will thus have total system losses on the order of $3 - 6~\text{dB}$ or higher, pushing the laser power requirement to at least $10 - 20~\text{W}$.

\subsection{Achieving laser power requirements}
As illustrated in the case study, practical long-range coherent lidars for vehicle applications require aggregate laser powers on the order of watts to tens of watts. These high power levels are best achieved using optical amplifiers. Although fiber amplifiers can readily reach these output powers, they tend to be too expensive for high volume vehicle applications. Semiconductor optical amplifiers (SOAs) currently appear to be the best option for achieving these output powers at reasonable costs.

Coherent lidars typically use photonic integrated circuits (PICs) to manage the transmission and detection of multiple lidar beams in parallel, and in many cases to also provide some degree of solid-state beam steering \cite{cvpoulton_ieeejstqe2019, crogers_nature2021, xzhang_nature2022, jkdoylend_sp2020, jkdoylend_apct2020}. Since coherent lidars only detect light in a single spatial optical mode, coherent lidar PICs make almost exclusive use of single-mode waveguides. The main challenge, therefore, is delivering watts to tens of watts of optical power into an array of single-mode PIC waveguides.

Although a wide variety of approaches for laser and SOA integration have been developed for PICs \cite{sshekhar_naturecomm2024, jystan_jlt2024}, they have typically been targeted at telecom applications which only require power levels in the milliwatts to tens of milliwatts. State of the art demonstrations of single-mode SOAs integrated with PICs have achieved approximately $100~\text{mW}$ per channel \cite{jkdoylend_sp2020, jkdoylend_apct2020, rkumar_jlt2024}. Reaching the aggregate powers required for long-range automotive lidar thus either requires tens to hundreds of SOA channels in parallel, implying high SOA integration yields and low cost per SOA channel, or significant increases in the output power of SOAs integrated with PICs.

\section{DISAMBIGUATING MULTIPLE RETURNS}
\label{sec:disambiguating_returns}
\begin{figure} [ht]
   \begin{center}
   \includegraphics[scale=0.75]{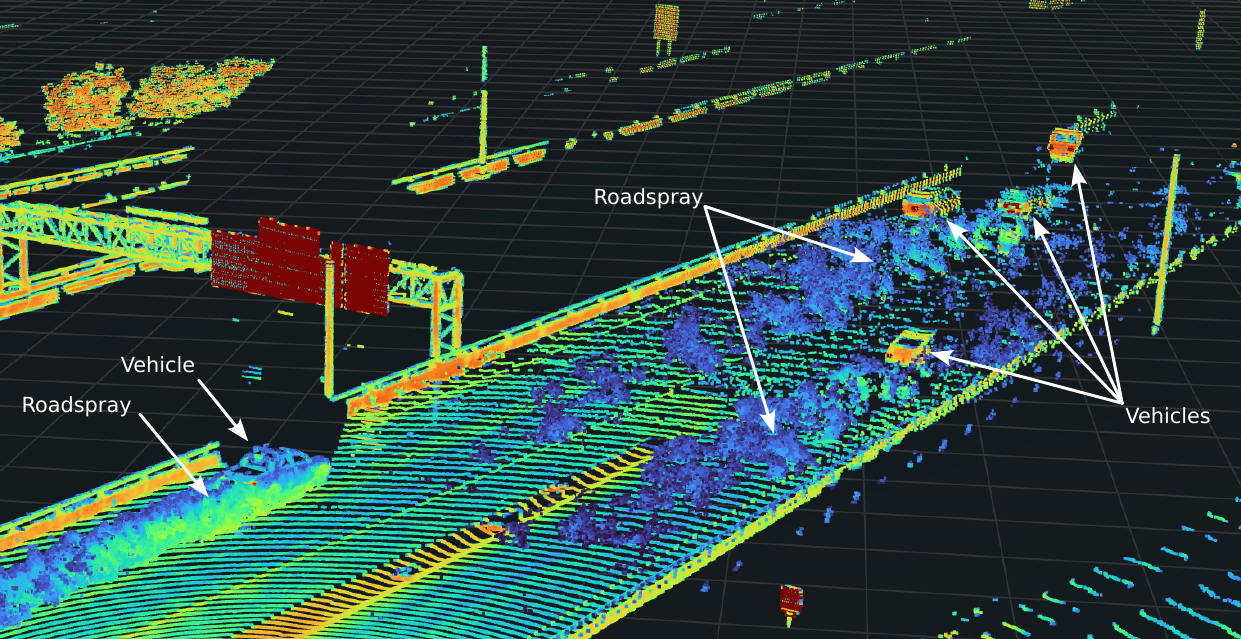}
   \end{center}
   \caption[example]{
        \label{fig:roadspray}
        Point cloud from a prototype time-of-flight Waymo lidar in wet weather with significant road spray. Courtesy of Waymo.
   }
   \end{figure}

Vehicle lidars operating in realistic driving environments frequently encounter multiple returns per shot. Weather is the dominant source of multiple returns, with fog, rain, snow, and roadspray all capable of generating relatively bright and dense lidar returns. A typical time-of-flight lidar point cloud in rainy weather is presented in figure \ref{fig:roadspray}, with dense plumes of road spray visible behind each vehicle in the scene. Static objects such as vegetation, wires, and chain link fences also frequently generate multiple returns. A lidar used as one of the primary sensors of an autonomous vehicle must be able to handle these multiple returns, penetrating through obscurants to see critical objects such vehicles and pedestrians.

\begin{figure} [ht]
   \begin{center}
   \includegraphics[scale=0.75]{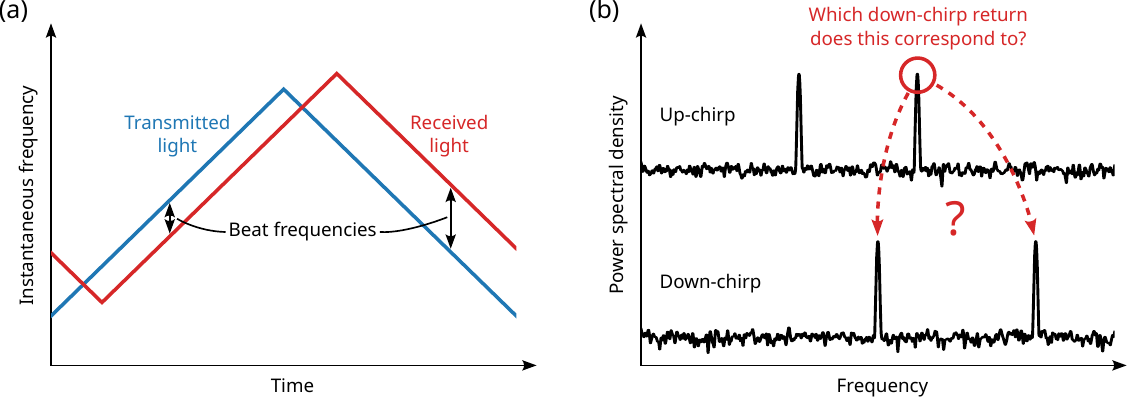}
   \end{center}
   \caption[example]{
        \label{fig:fmcw_disambiguation}
        Disambiguating multiple returns in a frequency-modulated continuous-wave (FMCW) lidar. (a) The canonical FMCW lidar scheme with a single return. Here, the transmit laser frequency is linearly swept up and down in a periodic fashion. The received light has both a time delay from the round-trip travel time, and a frequency shift due to the Doppler effect. By combining the transmitted and received light on a heterodyne detector and looking at the beat frequencies during the up- and down-chirps, one can calculate the position and velocity of the reflector. (b) If we have multiple returns, it is entirely unclear which detected tone from the up-chirp corresponds to which tone from the down-chirp.
   }
   \end{figure}

Unfortunately, the canonical frequency-modulated continuous wave (FMCW) scheme using two chirps cannot handle multiple returns. In this scheme, as illustrated in figure \ref{fig:fmcw_disambiguation}(a), the transmit laser frequency is linearly swept upwards and downwards in a periodic fashion at some ramp rate $\mu = df / dt$. The light received from the target has both a time delay proportional to the target distance, and a frequency shift proportional to the relative velocity of the target due to the Doppler effect. The beat frequencies $f_1$ and $f_2$ between the received and transmitted light during the up- and down-chirps respectively can be used to calculate the range and velocity of the target. The range $r$ is proportional to the mean of the beat frequencies, and is given by
\begin{align}
r = \frac{c \left(f_1 + f_2\right)}{4 \mu}.
\end{align}
Here, $c$ is the speed of light. Meanwhile, the velocity $v$ is proportional to the difference of the beat frequencies, and is given in terms of the laser wavelength $\lambda$ as
\begin{align}
v = \frac{\lambda \left(f_1 - f_2\right)}{4}.
\end{align}
This scheme does not generalize well to the case where there are returns from multiple targets. As illustrated in figure \ref{fig:fmcw_disambiguation}(b), if we have multiple returns, it is unclear which tone from the up-chirp corresponds to which tone from the down-chirp.

A number of more sophisticated modulation schemes can be used to handle multiple returns in coherent lidar. One simple solution is to use the FMCW scheme with 3 or more chirps \cite{mreiher_irs2008, mreiher_ieeerc2009a, mreiher_ieeerc2009b}. The additional chirps break the degeneracy present in the 2-chirp FMCW scheme, allowing arbitrary numbers of returns to be disambiguated except for rare corner cases. Phase-code schemes, where the phase of the transmit light is modulated by a psuedorandom binary sequence \cite{sbanzhaf_ieeetvt2021}, are another alternative.

\section{COMPUTATIONAL REQUIREMENTS}
Coherent lidar has heavy compute requirements when compared to time-of-flight lidars due to the need to compute fast Fourier transforms (FFTs) or cross-correlations. Application-specific integrated circuits (ASICs) optimized for this task are critical for controlling both costs and power consumption.

To illustrate this point, let us return to the example of the ADAS lidar we studied in section \ref{subsec:adas_lidar}. Given the maximum range of $250~\text{m}$ and a typical automotive lidar range resolution of $0.1~\text{m}$, the number of range bins $M$ is
\begin{align}
M = \frac{250~\text{m}}{0.1~\text{m} / \text{bin}} \approx 2048~\text{bins}.
\end{align}
The number of floating point operations (FLOPs) per FFT is approximately \cite{mfrigo_benchfft} given by
\begin{align}
5 \, M \log_2 M = 1.1 \times 10^5 ~\text{FLOPs} / \text{FFT}.
\end{align}
Meanwhile, the number of FFTs per second is given by
\begin{align}
\left(1.2 \times 10^7~\text{shots} / \text{second}\right)
\times \left(2~\text{chirps} / \text{shot}\right)
\times \left(4~\text{FFTs} / \text{chirp}\right)
= 9.6 \times 10^7~\text{FFTs} / \text{second}.
\end{align}
The total required compute is therefore
\begin{align}
\left(1.1 \times 10^5~\text{FLOPs} / \text{FFT}\right)
\times \left(9.6 \times 10^7~\text{FFTs} / \text{second} \right)
= 1.1 \times 10^{13}~\text{FLOPS}.
\end{align}
This is comparable to a high-end desktop graphical processing unit (GPU), which would add significant cost and power consumption to an automotive lidar. Hardware FFT engines optimized for this task would require significantly less chip area and power.

\section{CONCLUSIONS}
To summarize, a coherent lidar that is broadly adopted for ride-hailing autonomous vehicles would need to address the following challenges.
\begin{enumerate}
\item \textbf{Optimizing link transmission}: The optics design needs to carefully consider the unique needs of coherent lidar, namely narrow beams and limited angular scan rates.
\item \textbf{Laser power requirements}: It is necessary to achieve aggregate powers on the order of watts while maintaining narrow linewidths and single-mode beams.
\item \textbf{Disambiguating multiple returns}: Handling multiple returns is a critical requirement for any robust autonomous vehicle.
\item \textbf{Dedicated ASICs for compute}: The heavy compute requirements of coherent lidar can most easily be met in a cost- and power-efficient fashion by designing custom ASICs.
\end{enumerate}
Although these challenges have yet to be fully addressed in a cost-effective fashion, they are by no means insurmountable. Given the inherent advantages of coherent lidar, the rewards are high if these issues can be solved.

\appendix
\section{LINK TRANSMISSION MEASUREMENTS}
\subsection{Frequency-modulated continuous-wave (FMCW) lidar setup}
\label{sec:tlafmcw_setup}
\begin{figure}[h!t]
   \begin{center}
   \includegraphics[scale=0.75]{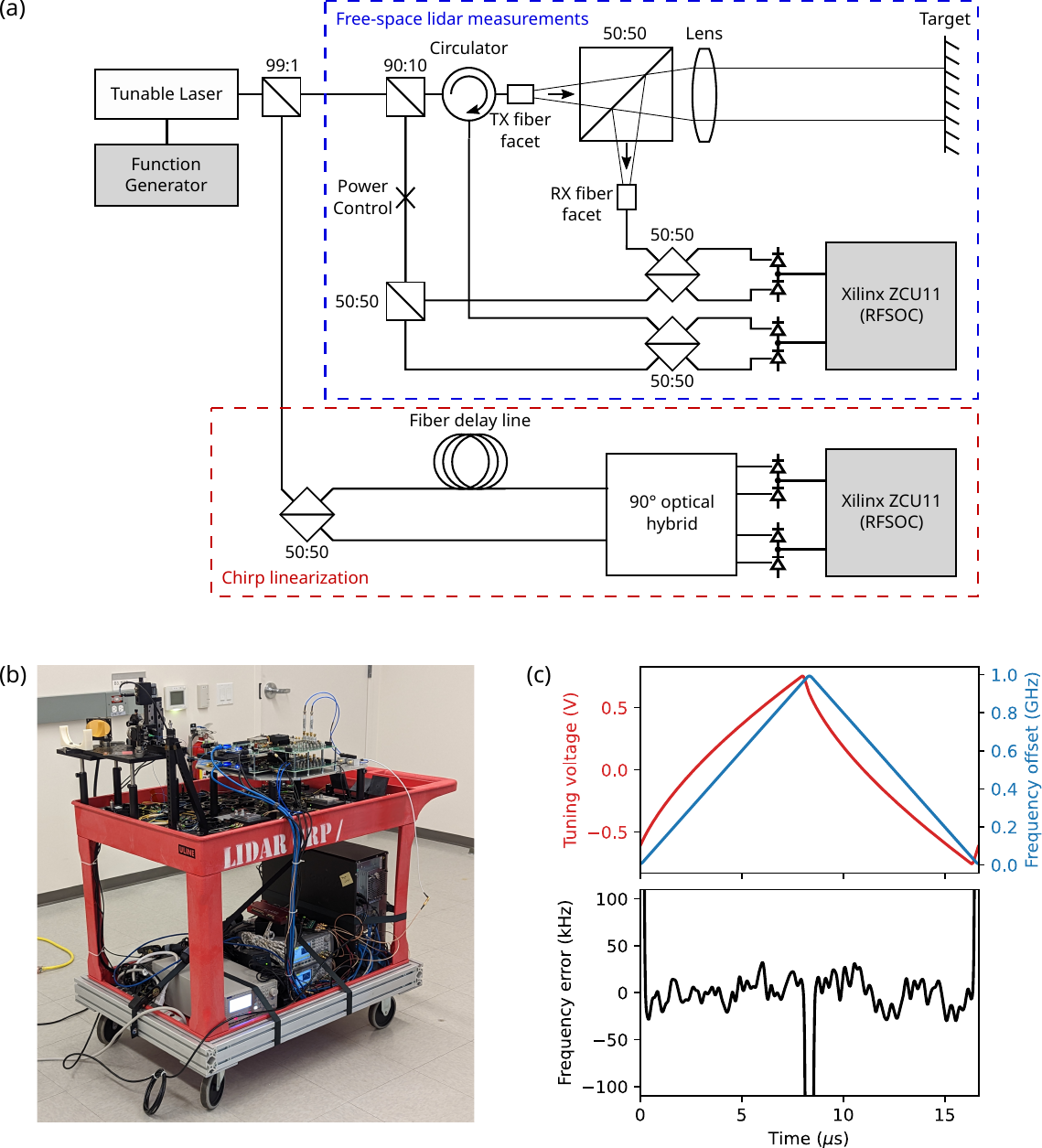}
   \end{center}
   \caption[example]{
        \label{fig:fmcw_lidar_setup}
        FMCW lidar setup. (a) The setup was built using off-the-shelf single-mode fiber optics. A distributed-feedback (DFB) laser was used as the light source, and was modulated to generate linear up- and down-chirps. One arm of the circuit, labeled ``chirp linearization'', was used to monitor the up- and down-chirp waveforms and provide the error signal for linearizing chirps. The other arm, labeled ``free-space lidar measurements'', was used as a free-space FMCW lidar. Only the monostatic receive path through the circulator was used for the measurements in this paper. (b) Photo of the FMCW lidar setup. (c) Up- and down-chirp waveforms after chirp linearization. The laser tuning waveform provided to the Koheron CTL200-2 laser driver board, and the resulting laser frequency waveform, is shown in the upper plot. The lower plot shows the error between the desired and actual laser frequency waveforms.
   }
   \end{figure}

The frequency-modulated continuous-wave (FMCW) lidar setup used for link transmission measurements is illustrated in figure \ref{fig:fmcw_lidar_setup}(a) and (b). The system was built around a distributed-feedback (DFB) laser operating at $1550~\text{nm}$ (Applied Optics DFB-1550-BF-50-CW-F2-N420). Xilinx ZCU111 RFSoC (radio-frequency system-on-a-chip) evaluation boards were used for data acquisition, and overall system control was handled by a workstation. 

Up- and down-chirps were generated by modulating the laser drive current to tune the laser frequency, as shown in figure \ref{fig:fmcw_lidar_setup}(c). Based on the frequency response of the laser tuning, the wavelength tuning mechanism is believed to be thermal: that is, modulating the drive current in turn modulates the power dissipated within the laser junction, causing the temperature of the laser waveguide to rise and fall with the drive current. To produce linear chirps, the instantaneous frequency of the laser was monitored using a Mach-Zehnder interferometer with a fiber delay line on one arm. The laser tuning waveform was then iteratively updated to minimize the error between the desired and actual chirp waveforms \cite{xzhang_oe2019}. The system consistently achieved frequency errors of $<100~\text{kHz}$ for $1~\text{GHz}$ bandwidth chirps, corresponding to a non-linearity of better than 1 part in $10^4$.

\subsection{Measuring link transmission}
\label{sec:link_measurement}
\begin{figure} [ht]
   \begin{center}
   \includegraphics[scale=0.75]{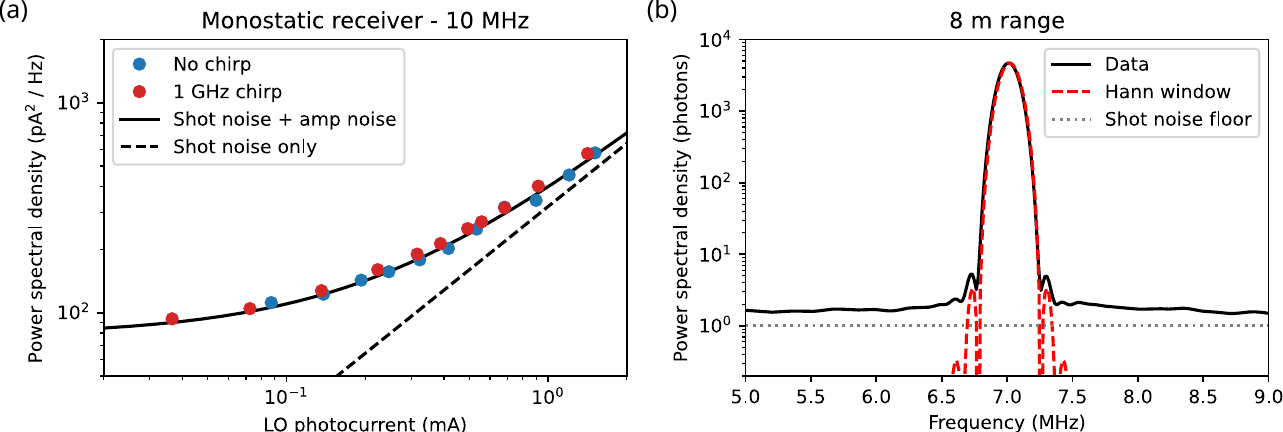}
   \end{center}
   \caption[example]{
        \label{fig:photon_counting_with_cnr}
        Photon counting using the carrier-to-noise ratio (CNR) for link transmission measurements. (a) The shot noise level at the receiver was calibrated before taking link transmission measurements. The noise as a function of local oscillator power is well described by a combination of shot noise and amplifier noise. (b) Power spectral density of a received lidar signal for a target at 8 meters. The lineshape closely matches the ideal lineshape of the FFT window. Due to the phase noise skirt of a strong zero-distance return from leakage in the optical circulator \cite{svenkatesh_jlwt1993}, the observed noise floor is somewhat higher than the shot noise floor.
   }
   \end{figure}

For the link transmission measurements, we used 1 GHz bandwidth, $8.3~\mu\text{s}$ long up- and down-chirps. As illustrated in figure \ref{fig:photon_counting_with_cnr}, the carrier-to-noise ratio (CNR) with respect to the shot noise level, which according to equation \ref{eqn:coherent_lidar_cnr} is equal to the number of received photons, was used to measure collection efficiency. The link transmission measurements were corrected for a variety of loss mechanisms in the setup, as listed in the table below. A Hann FFT window was used instead of a rectangular window to minimize the spectral sidelobes of bright returns, which would otherwise obscure dim returns.

\begin{center}
\begin{tabular}{ l c l }
  \toprule
  \textbf{Loss source} & \textbf{Transmission} & \textbf{Comments} \\
  \midrule
  50-50 beamsplitter, lens, mirror   & 0.403  & Measured with laser and power meter \\
  Fiber circulator and 50-50 coupler & 0.549  & Measured with laser and power meter \\
  Photodiode quantum efficiency      & 0.760  & From the Thorlabs PDC480C-AC user manual \\
  Processing loss                    & 0.667  & From the Hann FFT window \\
  \midrule
  \textbf{Total} & 0.112 & \\
  \bottomrule
\end{tabular}
\end{center}
   
\acknowledgments
We would like to thank Michael R. Matthews and Bernard Fidric for their contributions to coherent lidar investigations at Waymo.

\bibliography{references} 
\bibliographystyle{spiebib} 

\end{document}